# Plain *s*-wave superconductivity in single-layer FeSe on SrTiO$_3$ probed by scanning tunneling microscopy

Q. Fan[1], W. H. Zhang[1], X. Liu[1], Y. J. Yan[1], M. Q. Ren[1], R. Peng[1], H. C. Xu[1], B. P. Xie[1, 2], J. P. Hu[3, 4], T. Zhang[1, 2*], D. L. Feng[1, 2*]

[1] State Key Laboratory of Surface Physics, Department of Physics, and Advanced Materials Laboratory, Fudan University, Shanghai 200433, China
[2] Collaborative Innovation Center of Advanced Microstructures, Fudan University, Shanghai 200433, China
[3] Beijing National Laboratory for Condensed Matter Physics, Institute of Physics, Chinese Academy of Sciences, Beijing 100080, China
[4] Department of Physics, Purdue University, West Lafayette, Indiana 47907, USA

*E-mail of T. Zhang: tzhang18@fudan.edu.cn
*E-mail of D. L. Feng: dlfeng@fudan.edu.cn

**Single-layer FeSe film on SrTiO$_3$ (001) was recently found to be the champion of interfacial superconducting systems, with a much enhanced superconductivity than the bulk iron-based superconductors. Its superconducting mechanism is of great interest. Although the film has a simple Fermi surface topology, its pairing symmetry is unsettled. Here by using low-temperature scanning tunneling microscopy (STM), we systematically investigated the superconductivity of single-layer FeSe/SrTiO$_3$(001) films. We observed fully gapped tunneling spectrum and magnetic vortex lattice in the film. Quasi-particle interference (QPI) patterns reveal scatterings between and within the electron pockets, and put constraints on possible pairing symmetries. By introducing impurity atoms onto the sample, we show that the magnetic impurities (Cr, Mn) can locally suppress the superconductivity but the non-magnetic impurities (Zn, Ag and K) cannot. Our results indicate that single-layer FeSe/ SrTiO$_3$ has a plain *s*-wave paring symmetry whose order parameter has the same phase on all Fermi surface sections.**

Recently the discovery of enhanced superconductivity in single-layer FeSe on SrTiO$_3$(001) has attracted tremendous interest [1-9], not only for the new possible superconducting transition temperature records of Fe-based superconductors and interfacial superconductors (65K [3, 4] or even higher [9]), but also its intriguing mechanism that enhances the paring. Thus it is of great importance to understand the pairing symmetry and underlying electron structure of single-layer FeSe/SrTiO$_3$(001). Angle-resolved photoemission spectroscopy (ARPES) revealed that such films have only electron Fermi surfaces, similar to that of the alkali metal intercalated iron selenides (A$_x$Fe$_{2-y}$Se$_2$, A=K, Cs…) [3-5]. This seriously challenges the original

$s_\pm$-pairing scenario proposed for the iron pnictides that relies on the coupling between the electron pockets and the hole pockets at the Brillouin zone center [10,11]. Meanwhile both ARPES and previous STM studies found fully gapped superconducting state in single-layer FeSe, indicative of the absence of gap nodes [1, 3-5]. Various possible paring symmetries have been proposed for such systems with only electron pockets [12-19], such as plain *s*-wave paring [12-14], "quasi-nodeless" *d*-wave paring [15,16], and several new types of $s_\pm$ paring that involve the "folding" of Brillouin zone and band hybridization [17], orbital dependent pairing [18], or mixing of the even and odd-parity pairing [19]. Except the plain *s*-wave paring, all the other proposed pairing symmetries involve sign changing of the order parameter on different sections of the Fermi surface. To distinguish these scenarios, phase sensitive measurements are required, plus the detailed knowledge on the superconducting gap.

STM has been shown to be able to provide information on the pairing symmetry by measuring local response of superconductivity to impurities (in-gap impurity states) [20-22] and through quasi-particle interference (QPI) patterns [23-25]. Here we report a low-temperature STM study on the surface structure, superconducting state and pairing symmetry of single-layer FeSe/SrTiO$_3$(001). The samples were grown with molecular beam epitaxy (MBE) method under ultrahigh vacuum. We observed fully developed superconducting gap and magnetic vortex lattice, which confirms the high quality of the sample. The domain boundaries in the film are found to be no harm to the superconductivity. We then performed QPI mapping, where scatterings between and within the electron pockets are clearly present. The intensities of these scatterings as functions of energy near the superconducting gap edge, as well as their magnetic field dependence, are against the sign change between adjacent M points of the unfolded Brillouin zone. Meanwhile, by controllably introducing magnetic (Cr, Mn) and non-magnetic (Zn, Ag, K) atoms onto the sample surface as impurities, we found that Cr and Mn impurities induce pronounced bound states within the superconducting gap, but Zn, Ag and K impurities do not induce any of such states. These results evidence the phase-unchanging plain *s*-wave (or $s_{++}$ to differ from the $s_\pm$ one) pairing state in single-layer FeSe/SrTiO$_3$(001).

The experiment was conducted in a commercial STM (Unisoku) at the temperature of 4.2 K. The sample preparation and experimental details are described in the Methods section. Fig. 1a shows the typical topography of single-layer FeSe/SrTiO$_3$(001). The surface is atomically flat with winding, irregularly distributed domain boundaries. Fig. 1(b) shows a closer image of such a boundary. Weak 2×1 reconstructions can be observed within individual domains, and the reconstructions orient along different directions for adjacent domains (see also Fig. s1 in the Supplementary Material, the same notation is used below). One finds that the atom lattice extends continuously across the boundary. This is different from the previous study (ref. 1), in which the domains are separated by discontinuous boundaries. The difference may be due to some subtle change of post-annealing condition or substrate preparation, which is to be investigated. Detailed analysis shows that there is 1/2 unit cell offset along the [110] (Fe-Fe) direction between adjacent domains (Fig. s2), and the lattice is locally compressed at the domain boundary. The surface structure observed here is consistent with the model proposed in ref. 26, which ascribed the formation of 2×1 reconstruction and domains to ordered oxygen vacancies in SrTiO$_3$.

The typical superconducting gap spectrum of the film is shown in Fig. 1c, which is measured at 4K within a single domain. A fully gapped structure with double peaks at ±10 mV and ±15 mV is observed. The gap bottom with nearly zero tunneling conduction is of 5 meV wide. The spatial dependence of the superconducting gap is shown in Fig. 1d. A series of spectra are measured along a line cut of 20 nm long across the domain boundary (Fig. 1c). Within one domain, the spectra keep the same structure, although the coherence peaks vary slightly from point to point. At the domain boundaries, the gap still keeps the same width and the coherence peaks at ±10mV seem to be enhanced. Usually domain boundaries can be treated as line defects (impurities) that are non-magnetic in most cases. The preserved superconducting gap at the domain boundaries is consistent with the insensitivity of the superconductivity to non-magnetic impurities shown later in the remaining paper.

The superconductivity of the film is further investigated by imaging magnetic vortices. Fig. 2a shows a zero bias conductance (ZBC) mapping of a 70×70 nm$^2$ area, measured at B = 11T (perpendicular to the sample surface). The emergence of vortices is clearly reflected by the region with high ZBC values. Fig. 2b is a closer mapping of a single vortex. It is roughly diamond-shaped with the diagonal along the surface [100] (Se-Se) direction. This is in contrast with the elongated vortex observed in thick FeSe films [27], which implies the low energy states of single-layer FeSe are four-fold symmetric. Fig. 2d shows a series of spectra taken along a line cut across a vortex core. At the core center, a conductance peak near zero bias with slight asymmetry is observed. Such a vortex state is well expected for fully gapped superconductors, and it implies the film is in the clean regime [28]. Away from the core center, the vortex state splits into two symmetric peaks that eventually merge into the gap edges. Similar behavior has been observed in the *s*-wave superconductor NbSe$_2$ [28] and also in Fe-based superconductors like LiFeAs [29]. Fitting the line profile of Fig. 2b gives the Ginzburg-Landau coherence length ($\xi$) along the [100] and [110] directions, which are 3.18 nm and 2.45 nm, respectively (See Fig. 2c). The anisotropy of $\xi$ should relate to the anisotropy of the superconducting gap and the underlying Fermi surface [5].

To further examine the electronic structure of single-layer FeSe, we performed dI/dV mappings to reveal the QPI patterns. Here a clean 60×60 nm$^2$ area is chosen for the mapping (see Fig. s3). Figs. 3(a-f) show the fast Fourier transformation (FFT) of some of such maps taken at various energies (See Fig. s4 for more real space maps and FFTs). The FFTs are fourfold symmetrized since we didn't observe significant C4-C2 symmetry breaking on the QPI patterns (See Fig. s5). Ring-like scattering structures and the Bragg spots of the top Se lattice and Fe lattice can be observed in the FFTs. It is known QPI is caused by the scatterings between the states with the same energy. As observed by ARPES [3-5] and sketched in Fig. 3g, single-layer FeSe/SrTiO$_3$ in principle has two nearly circular electron pockets with finite ellipticity at each M point of the folded Brillouin zone (BZ). Since one of them is "folded" from the neighboring M points by the potential of Se lattice, it may have weak spectral weight and shown in dashed ellipse. At the energy outside of the superconducting gap, possible intra-pocket scattering vector ($q_1$) and that of inter-pocket scatterings ($q_2$) are illustrated in Fig. 3g. The reciprocal vector ($q_3$) that connects opposite M points is also marked. In Fig. 3h we show a simulated FFT by calculating the joint density of states: $JDOS(q) = \int I(k)I(k+q)d^2k$. Here we used the unfolded BZ and neglected the folded bands (The JDOS calculated by considering folded bands does not provide better simulation, see Fig. s6). One see that the basic features in Fig. 3(a-f) are mostly reproduced in the simulation. $q_2$ and $q_3$ coincide with the Bragg spots of the top Se lattice and the underlying Fe lattice, respectively. The center ring (referred as Ring 1 in the following text) is contributed by intra-pocket scatterings from all of the electron pockets, as exemplified by $q_1$. The rings centered at ($\pi$, $\pi$) (referred as Ring 2) and equivalent points are contributed by inter-pocket scatterings between the nearest neighboring electron pockets. The rings centered at (0, $2\pi$) (with weak intensity, referred as Ring 3) and equivalent points should have the same properties as Ring 1 since they just differ by a reciprocal vector ($q_3$). The anisotropy of the Ring 1 with broadenings around (0, 0)–(0, $2\pi$) directions could be due to the band ellipticity (which is 0.9 in the simulation). A feature not reproduced in simulation is that Ring 2 is actually consisted of arcs. It lacks of four-fold symmetry and does not show the similar anisotropy as Ring 1. This may require the consideration of different orbital characters of bands at neighboring M points (See Fig. s6 and its caption).

In Fig. 3i and 3j we show the line cuts extracted from the FFT images (in color scale), taken along (0, 0)–($\pi$, $\pi$) and (0, 0)–(0, $2\pi$) directions as marked in Fig. 3b. Except the fully gapped region, the dispersion of Ring 2 is clearly present in Fig. 3i. A parabolic fitting yields a band bottom at -60 mV, consistent with the previous ARPES data [2-4]. Ring 1 basically has a similar dispersion along (0, 0)–(0, $2\pi$) (Fig. 3j). But one found that in the range between 15 mV to 5mV, there are some additional features besides the parabolic fitting (marked by dashed curves in Fig. 3j). It can also be seen that the Ring 1 in Fig. 3c (measured at 9 mV) has different structure from the ones in Fig. 3a~3b measured above 15meV (More detailed energy dependence can be found in Fig. s4). Because this energy range is near superconducting gap edge, these features could be due to some gap anisotropy (as observed in the single-layer FeSe with more stretched lattice [5] and in LiFeAs [30]), and/or relate to the double-gap structure of the spectrum, which requires further study.

Next we focus on the investigation of the gap symmetry. Up to now various pairing scenarios have been proposed for the iron-based superconductors with only electron pockets [10-19]. For plain $s$ or $s_{++}$ pairing, the superconducting gap $\Delta_k$ has the same phase on all the electron pockets (Fig. 3g) [12-14]. For the "nodeless" $d$-wave pairing symmetry [15,16] (Fig. s6(a)), in the unfolded BZ, $\Delta_k$ changes sign between the nearest M points but is still nodeless. Normally, the finite folding potential will mix the Fermi pockets of different signs in the folded BZ, gap nodes will develop. However, it was argued that if the folding potential is not significant [11], a full gapped spectrum may be observed in this scenario. Meanwhile, various scenarios for $s_{\pm}$ pairing were proposed in the folded BZ [17-19] (see Fig. s6(b) and its caption). In these scenarios, $\Delta_k$ changes sign between the inner part and the outer parts of the two electron pockets at each M point.

QPI can provide information on the gap structure and symmetry. It is known when the energy is near or within the superconducting gap edge, the scattering is dominated by the Bogoliubov quasi-particles. Due to the coherence factors in the scattering matrix element [25], in the presence of non-magnetic scattering potentials, which are even under time-reversal, the scattering that preserves the sign of $\Delta_k$ will be suppressed, and the scattering that changes the sign of $\Delta_k$ will be enhanced [25,31,32]. In our case, the scatterers that generate the QPI at zero field (Fig. 3a-3f and Fig. s4) are likely from interface disorders/defects, which are expected to be mostly non-magnetic. Thus when the energy approaches the superconducting gap edge, the intensity of different scatterings will vary according to the pairing symmetry [32]. For example, in the case of $d$-wave as illustrated in Fig. s7(a), Ring 2 is from inter-pocket sign-changing scattering and may have enhanced intensity, while those sign-preserving Ring 1 and Ring 3 may have reduced intensity. In Fig. 4b we show the integrated FFT intensities over the three scattering rings, as a function of energy, after excluding the regions near (0,0) and the Bragg spots in the integration (see Fig. 4a). The intensities are all normalized at -30 meV, where the intensity is much less affected by the coherence factor. One found that the intensities of all the scattering rings actually have similar energy dependence near the gap edge, which are peaked around ±10mV and suppressed inside. This uniform energy dependence implies that all these scattering channels do not have significant difference, suggesting that the sign of $\Delta_k$ does not depend on the M point at which it is located. Thus it is against the "nodeless" $d$-wave pairing and consistent with the plain $s$ or $s_\pm$ wave pairing symmetries. Here the $s_\pm$-wave pairing cannot be distinguished from the plain $s$-wave since the outer and inner parts of the two electron pockets at the M point of the folded BZ are not resolved in the QPI pattern (See Fig. s6 and Fig. s7b-c).

The phase information can also be deduced though magnetic field dependence of the QPI. Because the vortices will introduce magnetic scattering potentials, which are odd under time-reversal, they will enhance the sign-preserving scatterings and suppress the sign-changing ones [25,31]. In Fig. 4c, we show the difference of the scattering intensities between B = 11 T and B = 0 T, taken at 10.5 mV (see Fig. s8 and s9 for more comparisons between dI/dV maps and FFTs measured at B=11T and B = 0T). However, apparently an overall suppression is observed for all the scattering rings. This could be due to the argument that the supercurrent surrounding vortices will induce Doppler shift to the quasi-particle energy, which tend to suppress the interference from all kinds of scatterings [31]. We then compared the relative change of intensities of different scattering rings in the field, as the function of energy. As shown in Fig. 4d, again all the scattering rings show similar behaviors. The largest suppression happens at around ±10 meV. Thus the sign structure of $\Delta_k$ on different M pockets is also expected to be similar here, against the $d$-wave pairing.

To further determine the pairing symmetry, we probed impurity-induced effects [20-22]. It is known that the response of superconductivity to local impurities depends on the pairing symmetry and the characteristic of the impurities [20]. For the plain $s$-wave pairing, the non-magnetic impurities will not suppress superconductivity according to Anderson's theorem, whereas magnetic impurities are pair-breakers that can induce quasi-particle bound states within the superconducting gap [20,21]. For the phase-changing pairing symmetries, such as $d$-wave or $s_\pm$-wave pairing, both magnetic and non-magnetic impurities are pair-breakers, as observed in cuprates with Ni and Zn impurities [22]. This is because even the scalar scattering will mix the quasi-particles with different phases, which tends to wipe out superconducting gap. For iron-pnictide superconductors, the proposed $s_\pm$ pairing scenario with phase change between hole and electron Fermi surfaces is expected to be fragile to non-magnetic impurities [33,34], which is indeed observed in recent STM measurements [35,36]. Thus it is highly desirable and practical to study pairing symmetry of single-layer FeSe using impurity effect [37].

The very first and crucial step to study impurity effect is to introduce well-defined impurities into the sample and identify whether they are magnetic or non-magnetic. This has been proven a nontrivial task [35,36]. In Fig. 1, we have shown that domain boundaries, which are most likely non-magnetic impurities, do not induce gap suppressions. To perform more determinate measurements, we applied a more controllable way to introduce impurities, that is, to directly deposit pure metal elements onto the cold sample surface (<50 K), which gives impurities in the form of single adatoms. For comparison, two kinds of magnetic atoms (Cr and Mn) and three kinds of non-magnetic atoms (Zn, Ag and K) were deposited separately onto the sample. These atoms all appear as point protrusions in the topography image [Fig. 5(a-b) and Fig. 6(a-c), and see also Fig. s10 for large scale images]. Their adsorption sites are all identified to be the hollow site of the top Se lattice (as marked in the image), which is expected to be the most stable adsorption site. Since the original Fe-Se bond should be intact, these impurity atoms shall keep their magnetic/non-magnetic characteristic after adsorption. In Fig. 5(c-d), we show the local tunneling spectra on the magnetic atoms. One sees that on the site of magnetic impurity Cr, a pair of sharp peaks appears at ±3mV and the superconducting coherence peaks are greatly suppressed. This hallmarks the impurity induced in-gap states. As moving away from Cr, the impurity states are weakened and the superconducting gap gradually recovers. The spatial distribution of these in-gap states is shown In Fig. 5e (3 meV) and 5f (-3 mV). They clearly display spatially inversed intensities, which is a characteristic of the in-gap Bogliubov quasiparticles with anti-phase relation at positive and negative energies [20]. Similar to the Cr case, the spectrum on the Mn impurity site also shows pronounced in-gap states at ±5 mV (Fig. 5d), and the gap recovered at about 2.4 nm away from the impurity.

Figs. 6(d-e) show the spectra taken across the non-magnetic impurities Zn, Ag and K, respectively. One see that the superconducting gap keeps undisturbed at impurity sites and nearby, and no evidence of in-gap states is observed on all of these impurities, although the coherence peaks show some random variations due to the slight inhomogeneity of the film (Fig. 1d). Therefore, the above observations indicate that only magnetic impurities can locally induce in-gap states and suppress the superconductivity. This is strongly against the phase-changing pairing symmetry such as $d$-wave and $s_\pm$-wave, and supports the phase-unchanging $s$ or $s_{++}$ wave.

Therefore by combining the impurity effects and QPI measurements, we show that the pairing symmetry of single-layer FeSe/SrTiO$_3$ is a plain *s*-wave. This represents a critical step toward the understanding of the mechanism behind this remarkable interfacial superconducting system. Scenarios that give the plain *s*-wave pairing symmetry, should be further examined, including the recent proposed phonon-mediated pairing enhancement scenario [7], the orbital fluctuation mediated pairing [38], and the extended *s*-wave with a leading cos($k_x$)cos($k_y$) paring [12].

Note: upon finishing this work, we noticed another independent STM study on single-layer FeSe/SrTiO$_3$, which mainly focused on the high energy band structure, by D. Huang, *et al.*, posted in arXiv: 1503.04792.

## Methods:

The SrTiO$_3$ (001) substrates with 0.5% Nb doping were cleaned by direct heating at 1250K in ultra-high vacuum. Single-layer FeSe films were grown by co-deposition of high purity Se (99.999%) and Fe (99.995%) on the substrate holding at 670 K. The films were post-annealed to 800 K for 2~3 hours after growth. Impurity atoms such as Cr, Mn, Zn, Ag and K are deposited onto the surface separately at low temperatures (~50K). Normal PtIr tips were used and cleaned by e-beam heating before measurement. Topographic images are taken with constant current mode with the bias voltage ($V_b$) applied to the sample. The tunneling conductance *dI/dV* is collected by standard lock-in method with a modulation frequency of 973 Hz.

**Acknowledgements**
We thank Professor Xin-Gao Gong and Professor Dung-Hai Lee for useful discussions. This work is supported by the National Science Foundation of China, and National Basic Research Program of China (973 Program) under the grant No. 2012CB921402, No. 2011CBA00112, and No. 2011CB921802.

**Author contributions**
The sample growth and STM measurements were performed by F. Qin, W. H. Zhang and T. Zhang. The data analysis was performed by F. Qin, X. Liu, J. P. Hu, T. Zhang and D. L. Feng. T. Zhang and D. L. Feng coordinated the whole work and wrote the manuscript. All authors have discussed the results and the interpretation.

**Additional information**
Competing financial interests: The authors declare no competing financial interests.

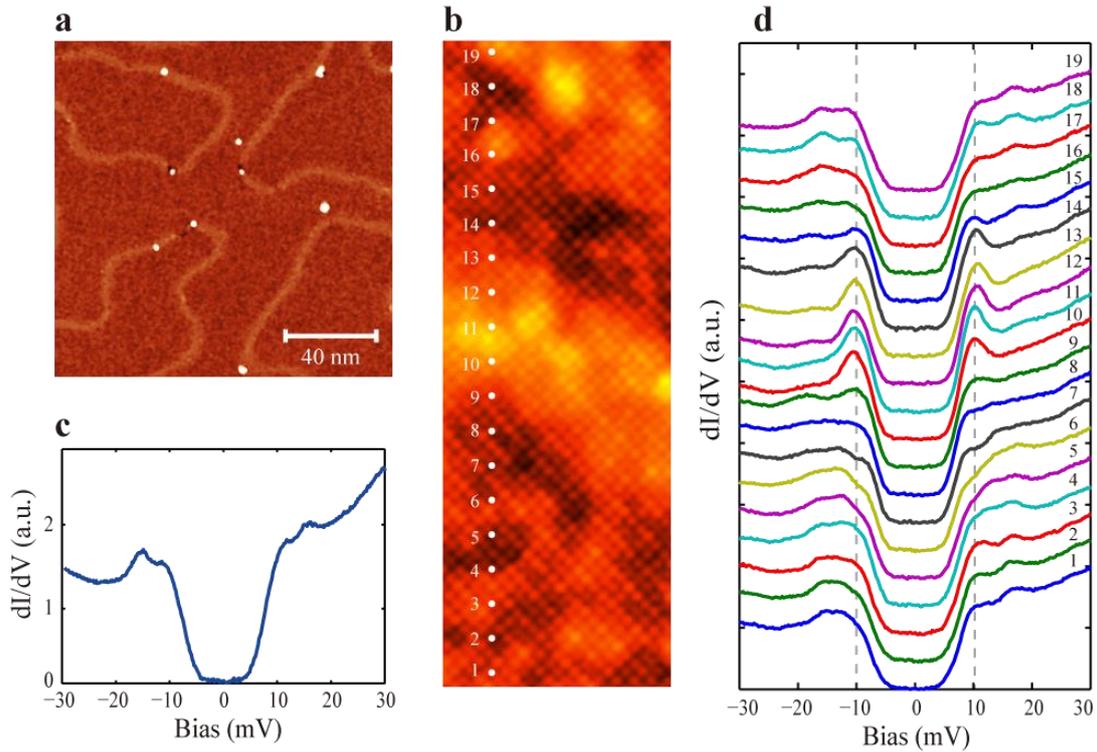

**Figure 1 | Surface topography and superconducting gap spectrum of single-layer FeSe/SrTiO$_3$(001).** **(a)** Large scale topographic image of the film (150×150 nm$^2$, $V_b$ = 3 V, I = 50 pA) **(b)** Topographic image around a domain boundary (20×9 nm$^2$, $V_b$ = 30 meV, I = 50 pA). The white spots and numbers indicate the locations where the spectra in panel **d** were taken. **(c)** A typical dI/dV spectrum taken on the film, shows the superconducting gap. **(d)** A series of spectra taken along a line crossing the domain boundary, as marked in panel **b**.

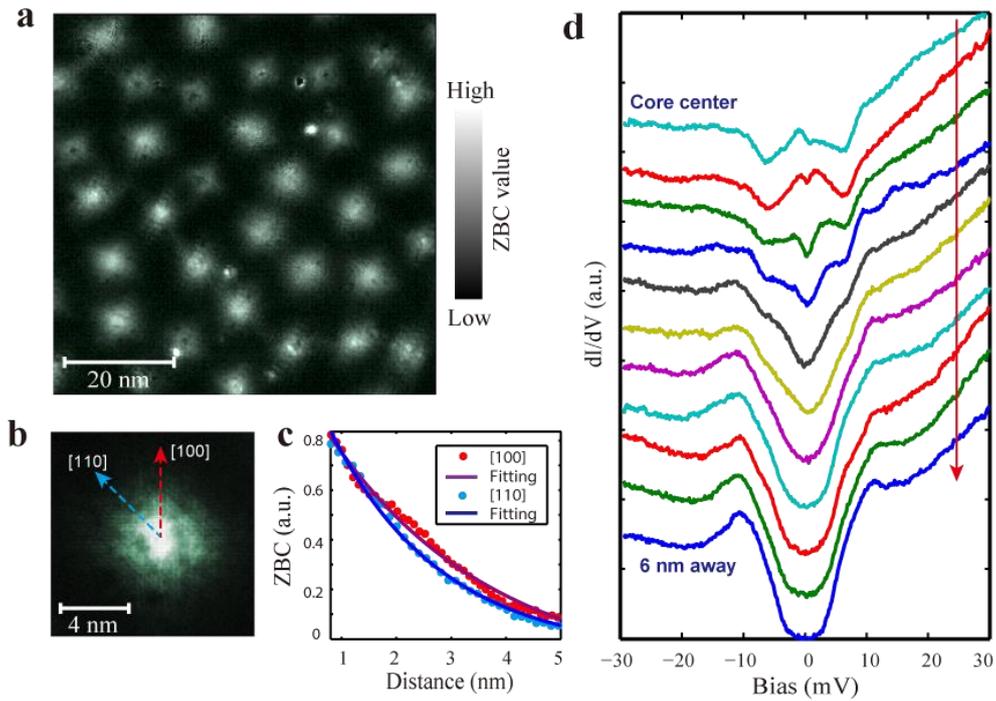

**Figure 2 | Magnetic vortex states of single-layer FeSe/SrTiO$_3$(001). (a)** A zero bias conductance (ZBC) mapping taken at B = 11T (70×70 nm$^2$), shows the emergence of vortex lattice. **(b)** A ZBC map around a single vortex. **(c)** Line cuts taken along the [100] and [110] directions in panel **b**, together with exponential fittings (y $\propto$ exp(-x/$\xi$)). **(d)** A series of spectra taken along the red arrow in panel **b**.

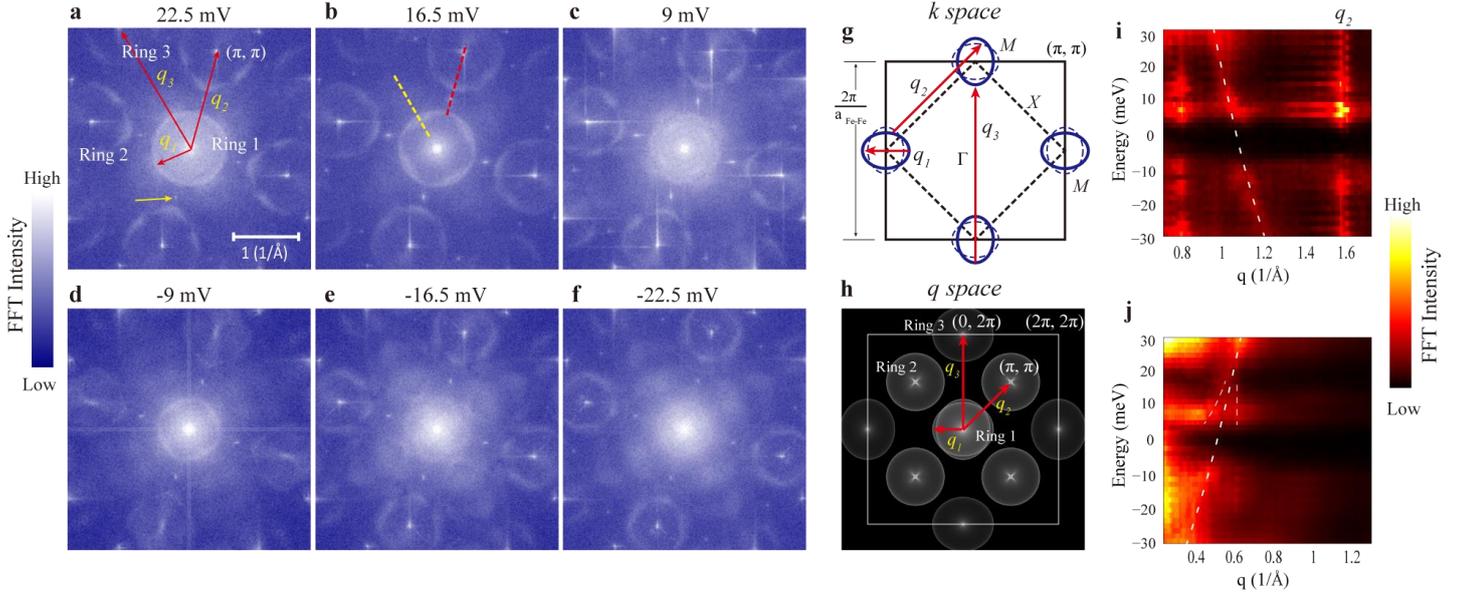

**Figure 3 | QPI mappings of single-layer FeSe/SrTiO$_3$(001). (a-f)** Fast Fourier transformation (FFT) of real space dI/dV maps taken at various bias (as labeled). The FFTs are four-fold symmetrized and shown in logarithmic scale. The positions of scattering vectors ($q_1$, $q_2$, $q_3$) are marked in (a). Note that $q_2$ and $q_3$ coincide with the Bragg spots of top Se lattice and Fe lattice, respectively. The yellow arrow in (a) points the spots of 2×1 reconstruction. **(g)** Schematic of the Brillouin zone (BZ) and Fermi surface of single-layer FeSe/SrTiO$_3$(001). Black solid square is unfolded BZ with the full width of (2π/a$_{Fe-Fe}$). The dashed square is folded BZ. Blue solid (dashed) ellipse are electron (folded) pockets at the M points. Possible scattering vectors ($q_1$, $q_2$, $q_3$) are marked by red arrows. **(h)** Simulated FFT image according to the electronic structure in the unfolded zone, at the energy outside of the superconducting gap. A band ellipticity of 0.9 is used. **(i)** and **(j)** The line cuts extracted from the FFTs along the red and yellow dashed line shown in panel **b**, respectively. The white dashed curves are parabolic fittings of the dispersive scattering rings. The short dashed lines indicates additional features at 15meV ~ 5meV.

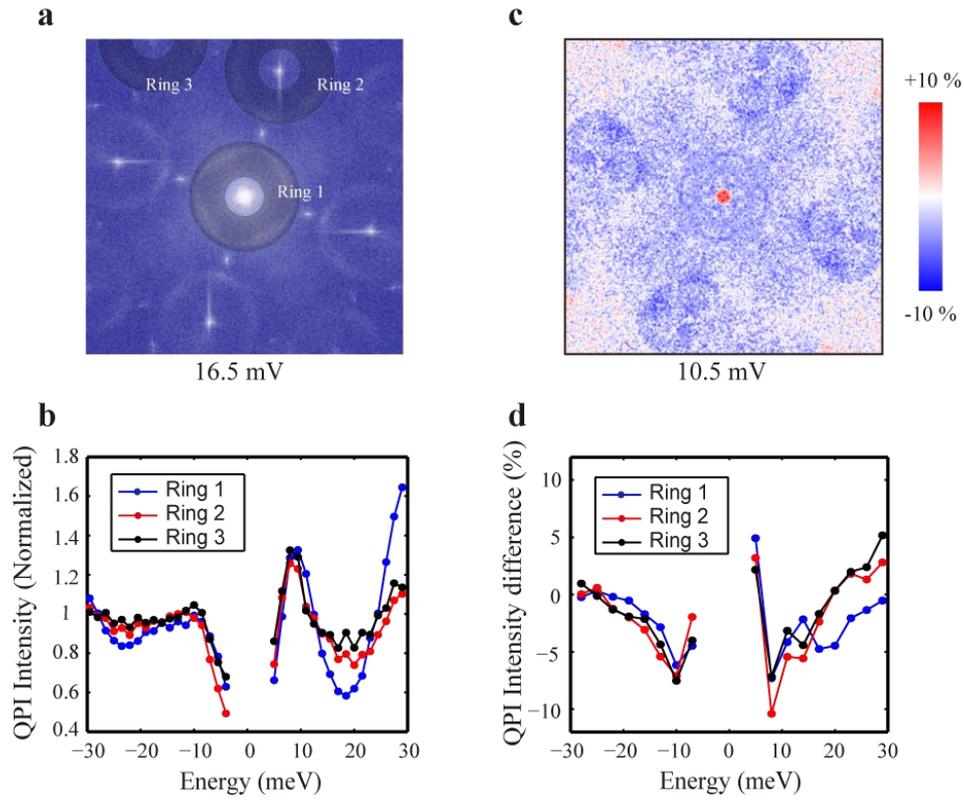

**Figure 4 | The QPI intensity analysis for different scattering channels.** (a) QPI pattern at 16.5 meV, where the shaded areas show the integration windows for different scattering rings. (b) Integrated intensity (normalized at -30 meV) of the three sets of scattering rings, as the function of energy. (c) The relative difference between the QPI maps at B = 11T and B = 0T, calculated by $(FFT_{11T} − FFT_{0T}) / (FFT_{11T} + FFT_{0T})$. (d) The relative difference of integrated intensity at B = 11T and B=0T as the function of energy for the three set of scattering rings. The data points in the fully gapped region (-5meV ~ 5 meV) are neglected in panels **b** and **d**

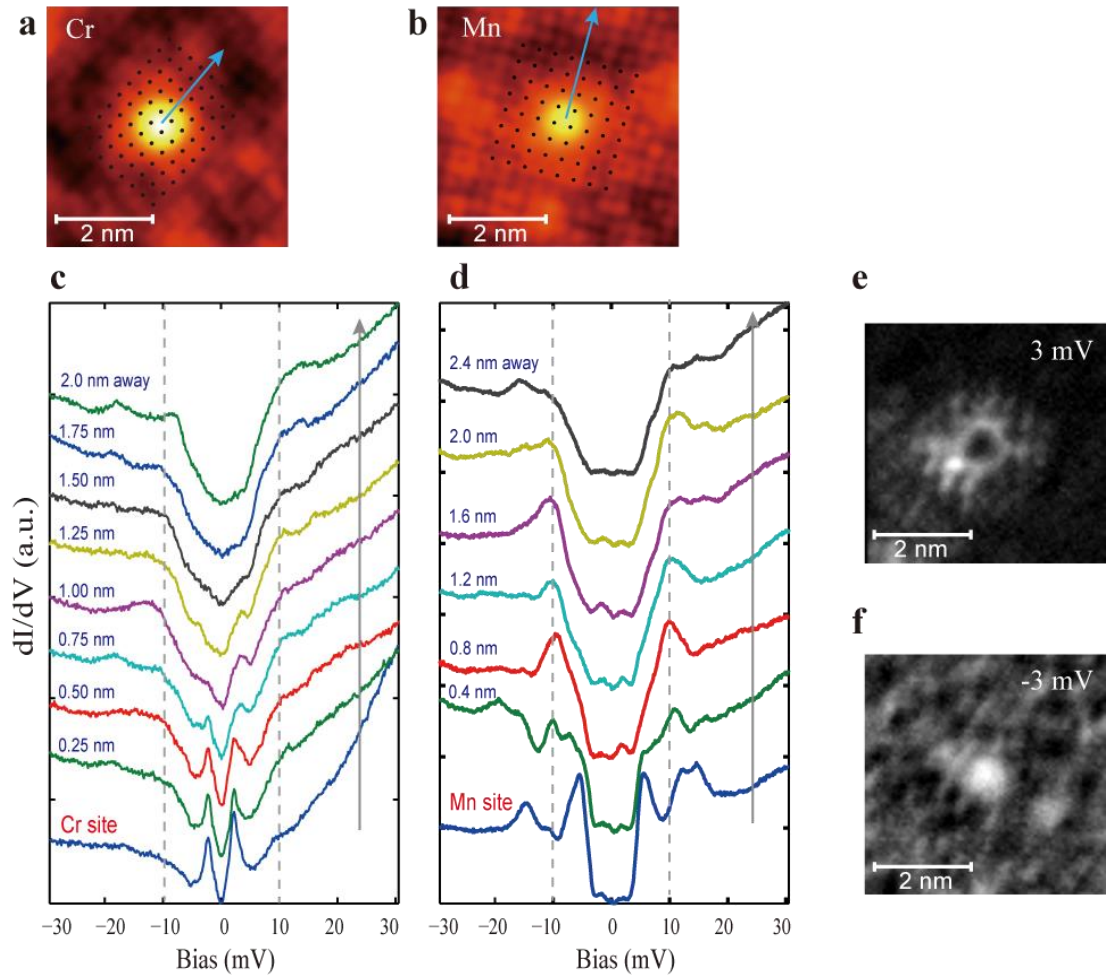

**Figure 5 | In-gap states induced by magnetic impurity Cr and Mn. (a-b)** Topographic images of Cr and Mn single adatoms (5×5 nm² size). The black spots represent the position of surface Se lattice. **(c)** and **(d):** Spectra taken along the arrows shown in **(a)** and **(b)**, respectively. The distances of the measuring points to the center of the atom are marked on the left. **(e)** and **(f)**: dI/dV mappings taken at 3 mV and -3 mV around a Cr atom, respectively. Mapping size is 5×5 nm².

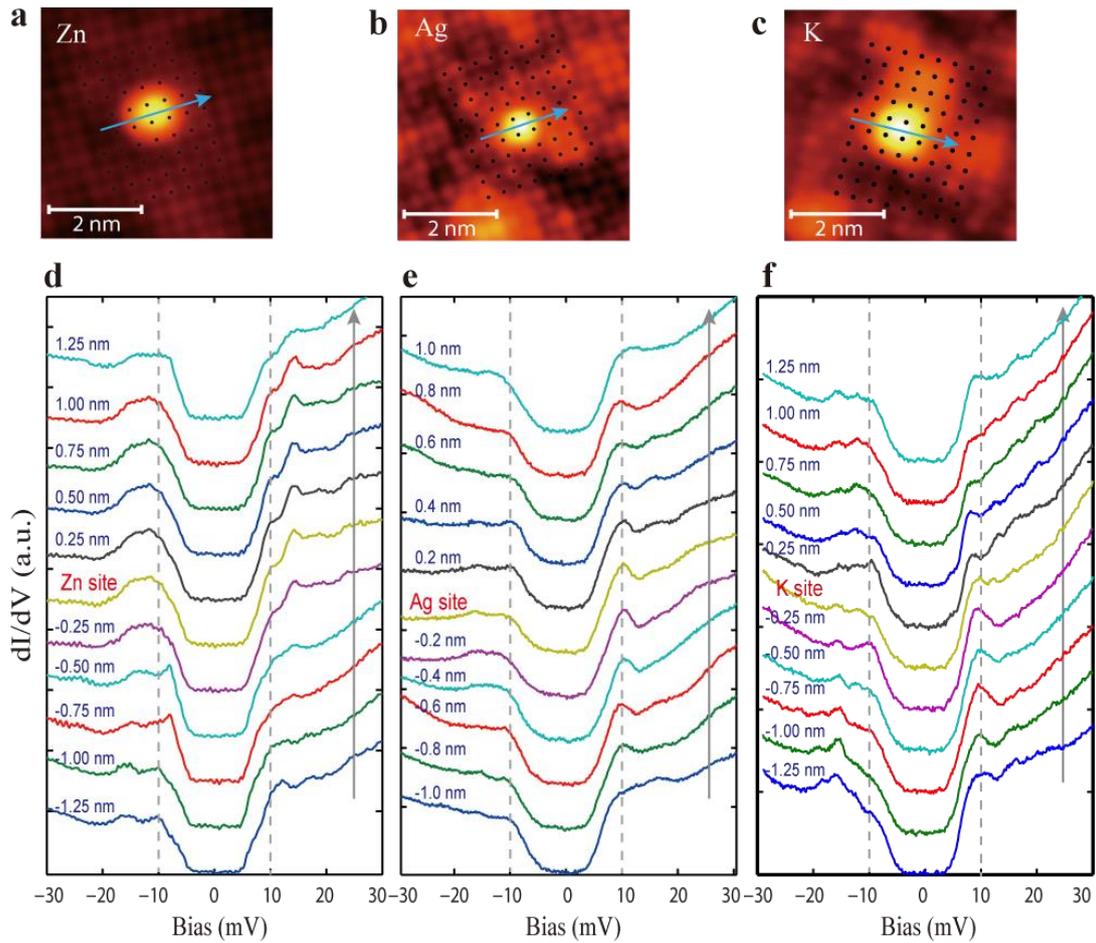

**Figure 6 | Absence of the in-gap states on non-magnetic impurities (Zn, Ag and K). (a-c)** Topographic image (5×5 nm² size) of Zn, Ag and K adatom, respectively. Black spots represent the position of surface Se lattice. **(d-f)** Spectra taken along the arrows shown in **(a-c)**, respectively. The distances of the measuring points to the center of the atom are marked on the left.

# Supplementary Materials for

# Plain *s*-wave superconductivity in single-layer FeSe on SrTiO$_3$ probed by scanning tunneling microscopy

Q. Fan[1], W. H. Zhang[1], X. Liu[1], Y. J. Yan[1], M. Q. Ren[1], R. Peng[1], H. C. Xu[1], B. P. Xie[1,2], J. P. Hu[3,4], T. Zhang[1,2*], D. L. Feng[1,2*]

[1] State Key Laboratory of Surface Physics, Department of Physics, and Advanced Materials Laboratory, Fudan University, Shanghai 200433, China
[2] Collaborative Innovation Center of Advanced Microstructures, Fudan University, Shanghai 200433, China
[3] Beijing National Laboratory for Condensed Matter Physics, Institute of Physics, Chinese Academy of Sciences, Beijing 100080, China
[4] Department of Physics, Purdue University, West Lafayette, Indiana 47907, USA

**I**. Surface domain structures of single-layer FeSe/SrTiO$_3$ (001)

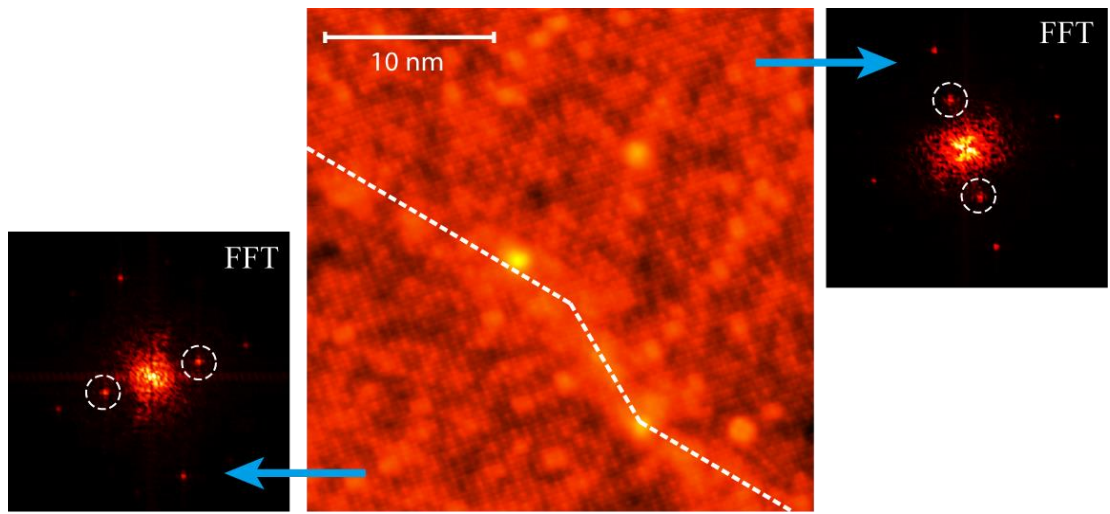

**Fig. s1 | Domains and 2×1 reconstructions.** Middle: A 30×30 nm$^2$ STM image shows two domains separated by a domain boundary (dashed line). Left and Right: Fast Fourier transformation (FFT) of the two different domains. Dashed circles mark the spots of 2×1 reconstructions. They have different orientations (perpendicular to each other) in different domains.

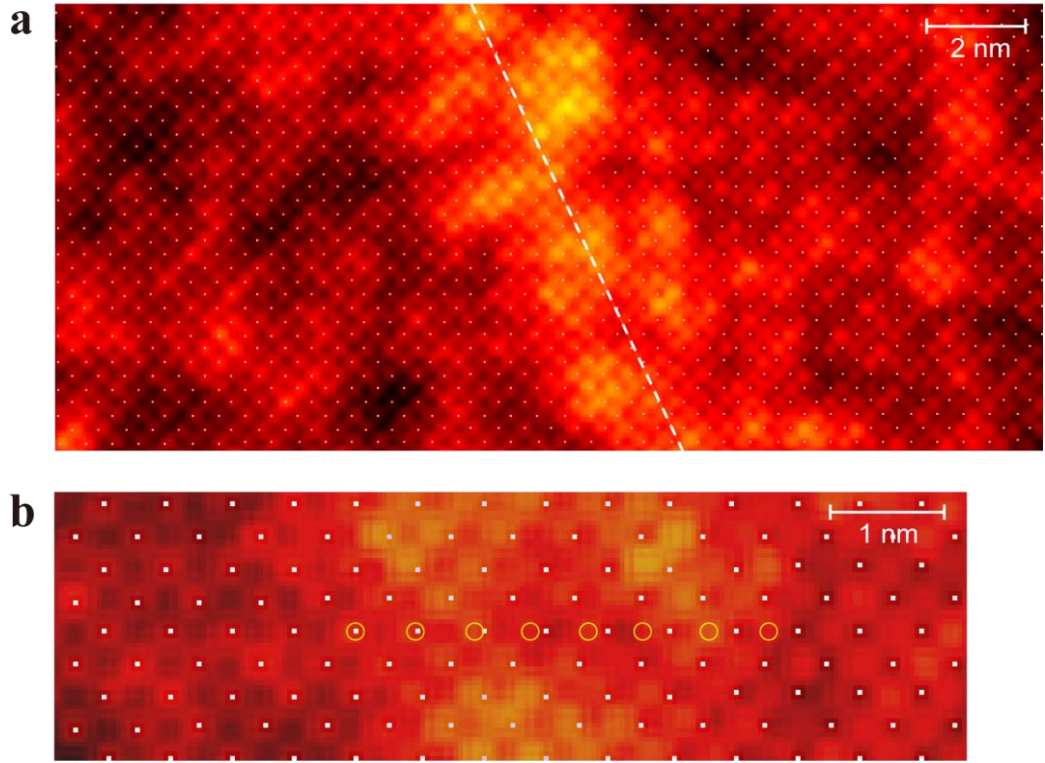

**Fig. s2 | The atom lattice around the domain boundary. (a)** A 20 nm ×9 nm STM image with a domain boundary in the middle (dashed line). A grid (white spots) is drawn to match the lattice of the left domain. After crossing the domain boundary, the grid mismatches the lattice of the domain on the right with 1/2 unit cell offset along the [110] (Fe-Fe) direction. This implies that there are two equivalent epitaxial sites of the single-layer FeSe on SrTiO$_3$ (001), agrees with model proposed in ref. 26. Similar result has been reported in Ref. s1, but here the lattice continues across the domain boundary. **(b)** A zoom-in image of the dashed region in panel **a**. By tracing the Se lattice (yellow circles), one found that the film is locally **compressed** by 1/2 unit along [110] at the domain boundary. The compressed region is about 3.5 nm wide.

## II. Additional images of QPI mappings

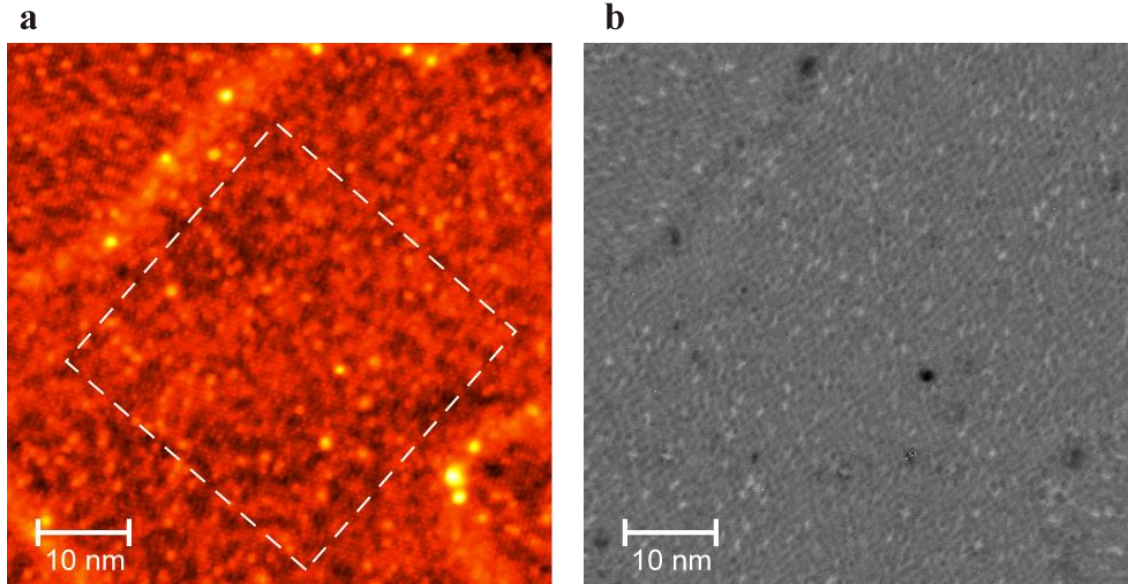

**Fig. s3 | (a)** Topographic image of the surface area for dI/dV mapping (size: 60 nm × 60 nm). Dashed square is a region within a single domain. The FFTs shown in Fig. s5 were performed in this region. **(b)** A typical dI/dV mapping (E = 24 meV) measured in panel **a**. Set point: $V_b$ = 30 mV, I = 100 pA, $\Delta V$ = 1 mV. The mapping has 430 × 430 pixels.

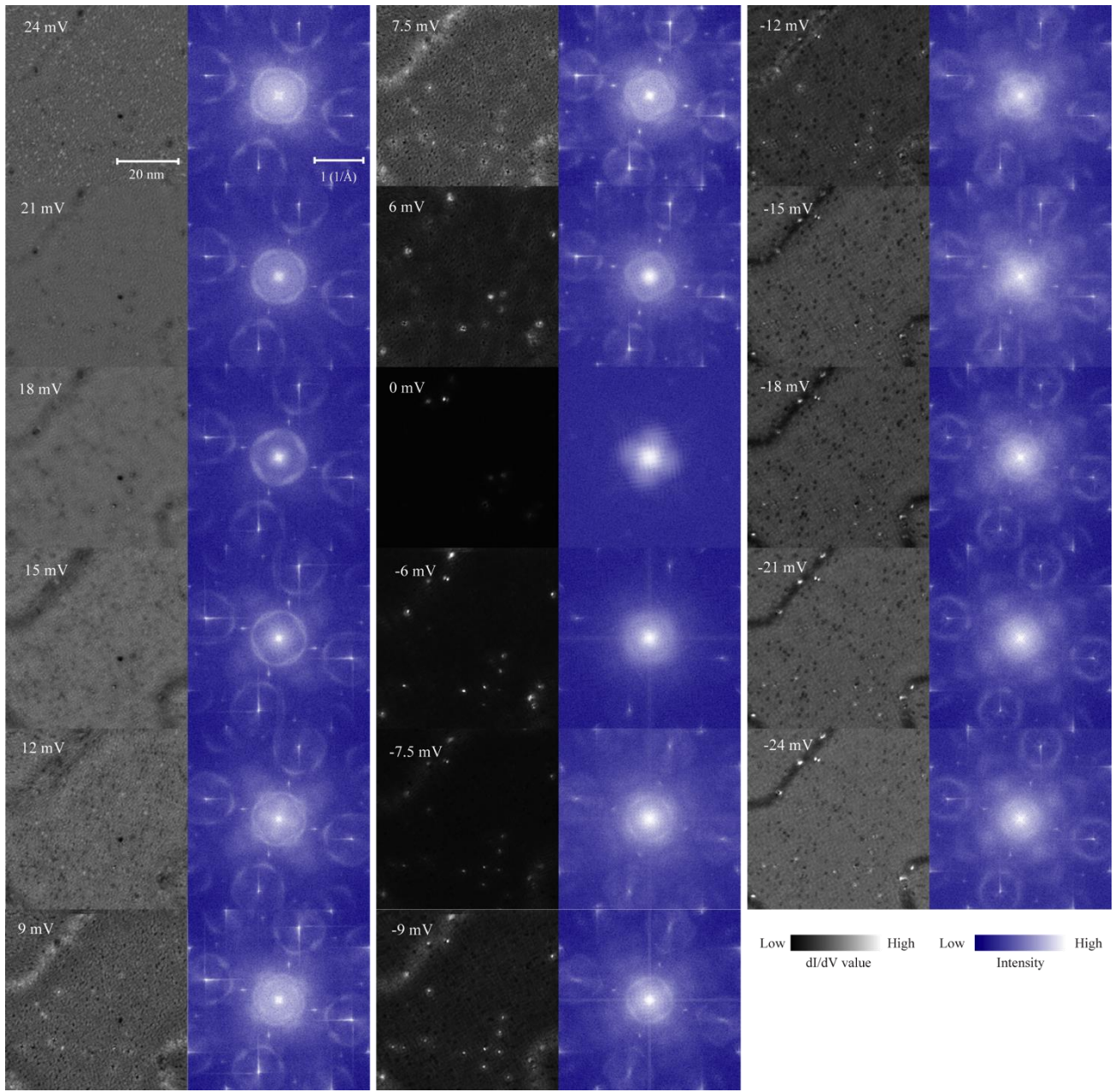

**Fig. s4** | The selected dI/dV maps taken in the area shown in Fig. s3 at various energies, and the corresponding FFT images (four-fold symmetrized). The mapping bias are labeled on the images. In the range of 15mV ~ 6mV, the shape of center scattering ring (Ring 1) show some subtle change, which may relate to gap anisotropy.

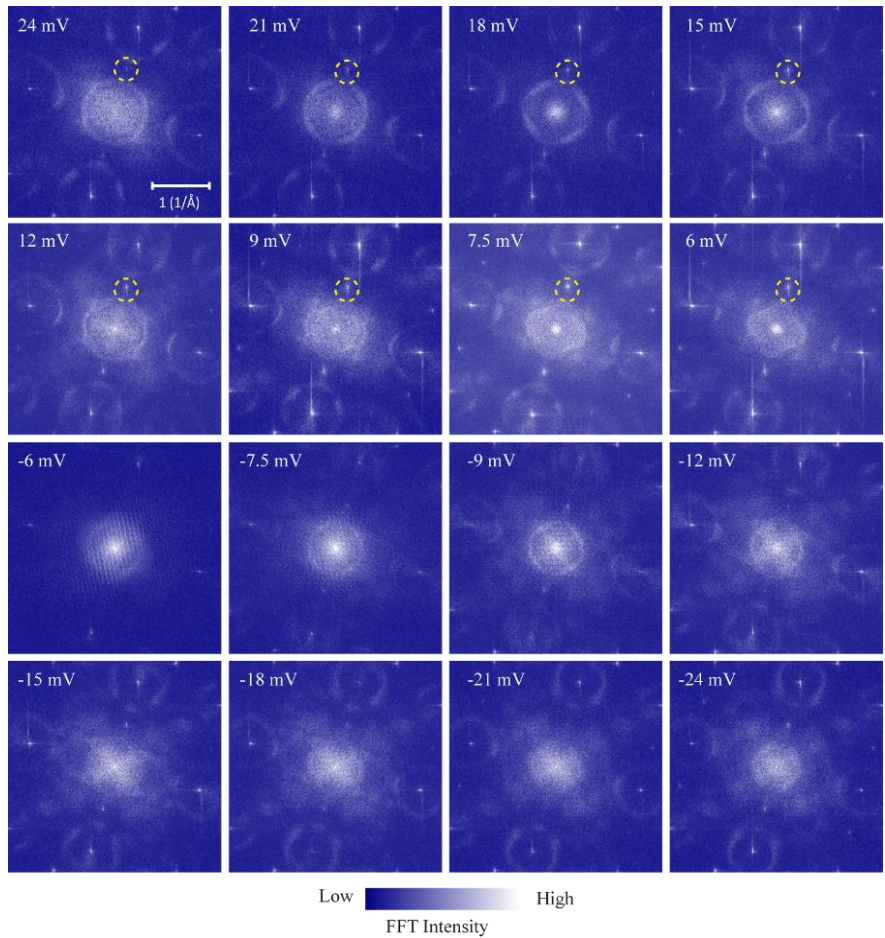

**Fig. s5 |** Raw FFT images (without symmetrization) of the **single domain region** marked by the dashed square in Fig. s3. The QPI patterns are still nearly four-fold symmetric, except the spots of 2×1 reconstruction (dashed circles). This is consistent with the four-fold symmetric vortex core observed in Fig. 2. The mapping biases are labeled on each images.

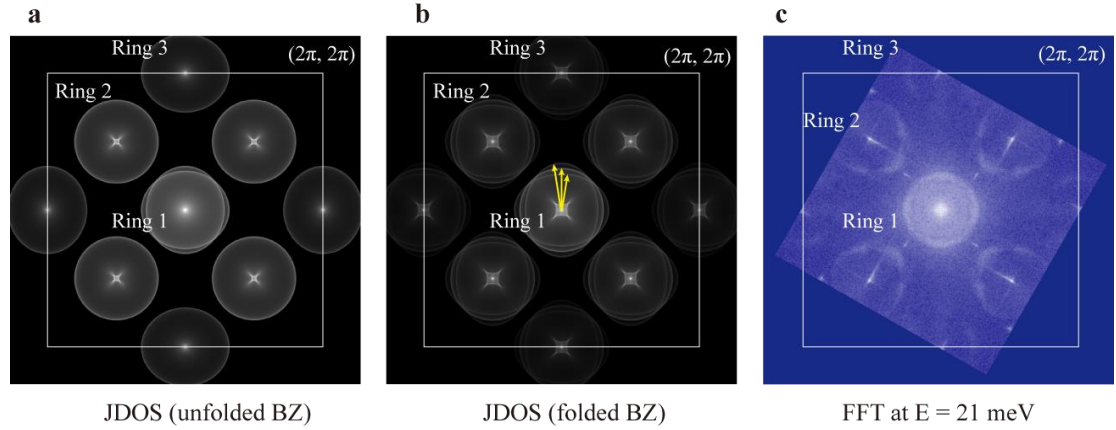

|   a   |   b   |   c   |
|:-----:|:-----:|:-----:|
| JDOS (unfolded BZ) | JDOS (folded BZ) | FFT at E = 21 meV |

**Fig. s6 | Simulated FFT patterns. (a)** JDOS calculated with unfolded BZ (neglecting the "folded" pockets). **(b)** JDOS calculated with folded BZ with two electron pockets (of the same spectra weight) at each M point. **(c)** QPI pattern at E = 21 meV, which is rotated for easy comparison with the simulated patterns. One found that the Ring 2 is circular in panel **a**, even upon considering the band ellipticity. However, the Ring 2 (also for Ring 3) in panel **b** has the same structure as Ring 1, which split into three sub-sets along (0, 0)-(0, 2π) directions (marked by the arrows in (b)). These features come from the folded bands. However, in all the experimental QPI patterns, such as the one in panel **c,** Ring 2 is actually consisted of arcs and lacks four-fold symmetry. Moreover, Ring 2 and Ring 1 do not have the similar anisotropy. These features are not reproduced either in panel **a** or in panel **b**, which implies the simulation may need to consider other factors like orbital structure of the neighboring M points. The three sub-sets of $q_1$ is not clearly resolved in all the QPI mappings, instead, the rings appear to have finite width. This may be due to the intrinsically weak BZ folding and/or limited k-resolution of the mapping.

# III. Additional information on the pairing symmetry and related QPI measurements.

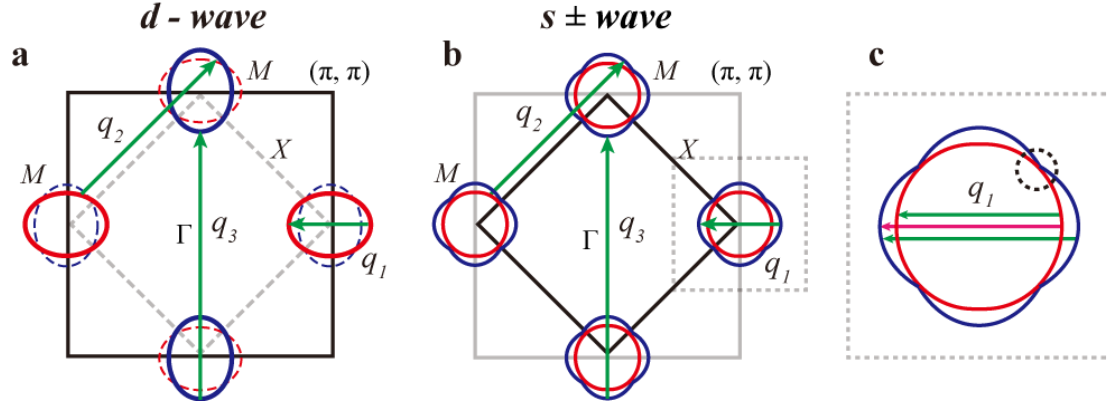

**Fig. s7 | Schematics of possible pairing symmetries of single-layer FeSe/SrTiO$_3$ (001) with sign-changing.** (**a**) "Quasi-nodeless" *d*-wave pairing state [15-16], the dashed ellipses show folded bands. It was argued that if the folded bands are weak, this pairing could be nodeless [11]. (**b**) $s_\pm$ wave pairing states [17-19] in the folded BZ case. The $\Delta_k$ on the red and blue parts in panel **a** and **b** are of opposite signs. (**c**) Detailed structure near the M point of the $s_\pm$ wave pairing scenarios. Note the scenario in ref. 17 requires band hybridization at the band crossing point (marked by dashed circle). Nevertheless, for all the $s_\pm$ wave pairing scenarios, the intra-pocket scattering $q_1$ should have three sub-sets, as shown by the three arrows in panel **c** and reflected in the JDOS simulation in Fig. s6(b). Two of these three sub-sets are sign-preserving (green arrows) and the other one is sign-changing (purple arrow). However, these sub-sets cannot be well resolved in the QPI patterns (Fig. s4 and Fig. s9), because of the intrinsically small BZ folding and/or limited k-resolution of the mapping. Thus just from QPI measurement here the $s_\pm$ wave pairing cannot be unambiguously distinguished from the $s_{++}$ wave paring.

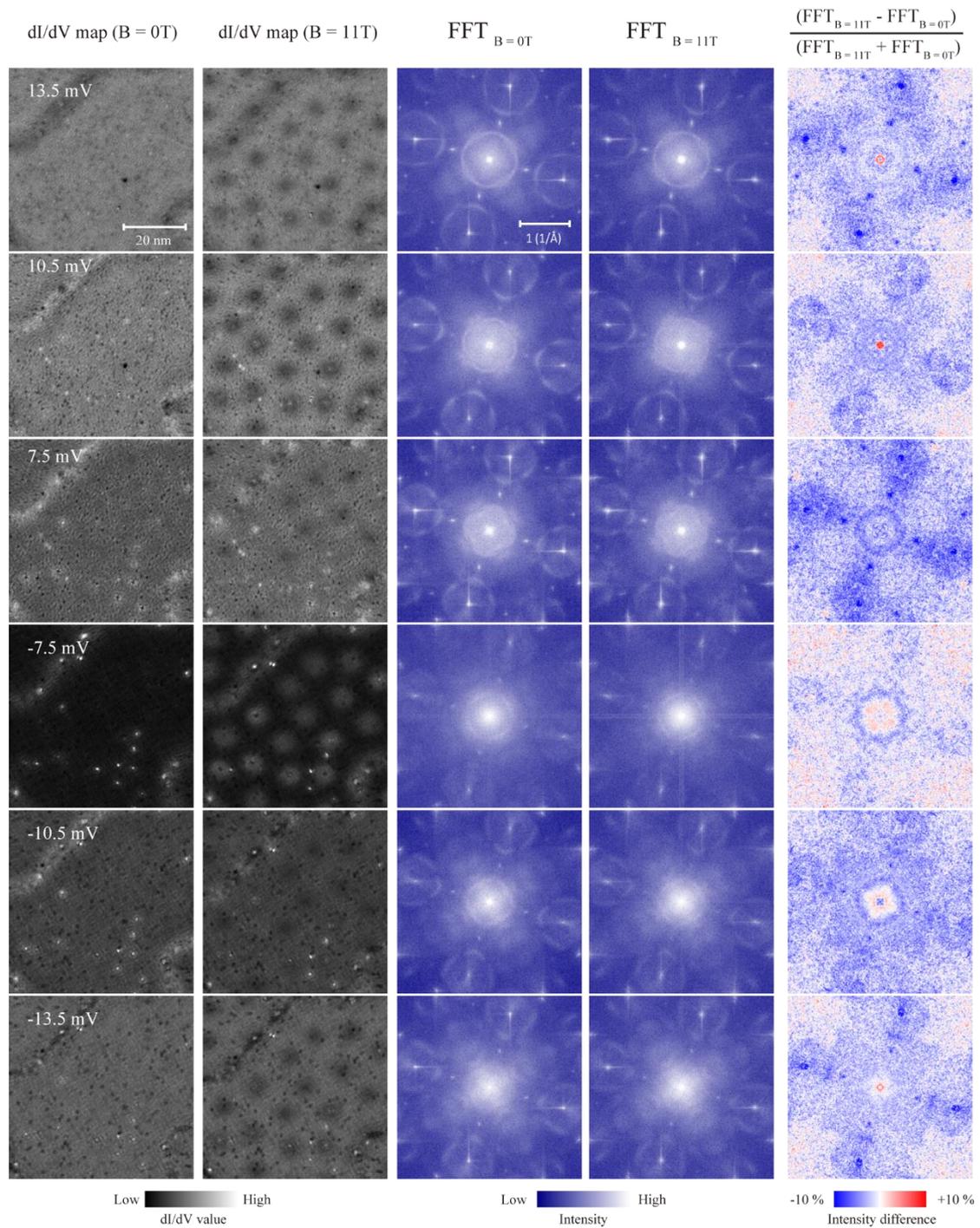

**Fig. s8 | Comparison of the dI/dV maps and their FFTs taken at B = 0T and B = 11T, at the energies close to the superconducting gap.** An overall suppression of the scatterings can been seen in the last column, which is calculated by $(FFT_{11T} - FFT_{0T}) / (FFT_{11T} + FFT_{0T})$. All the dI/dV maps are taken at the same set point of $V_b = 30$ mV, $I = 100$ pA, $\Delta V = 1$ mV. The mapping bias are labeled on the first column.

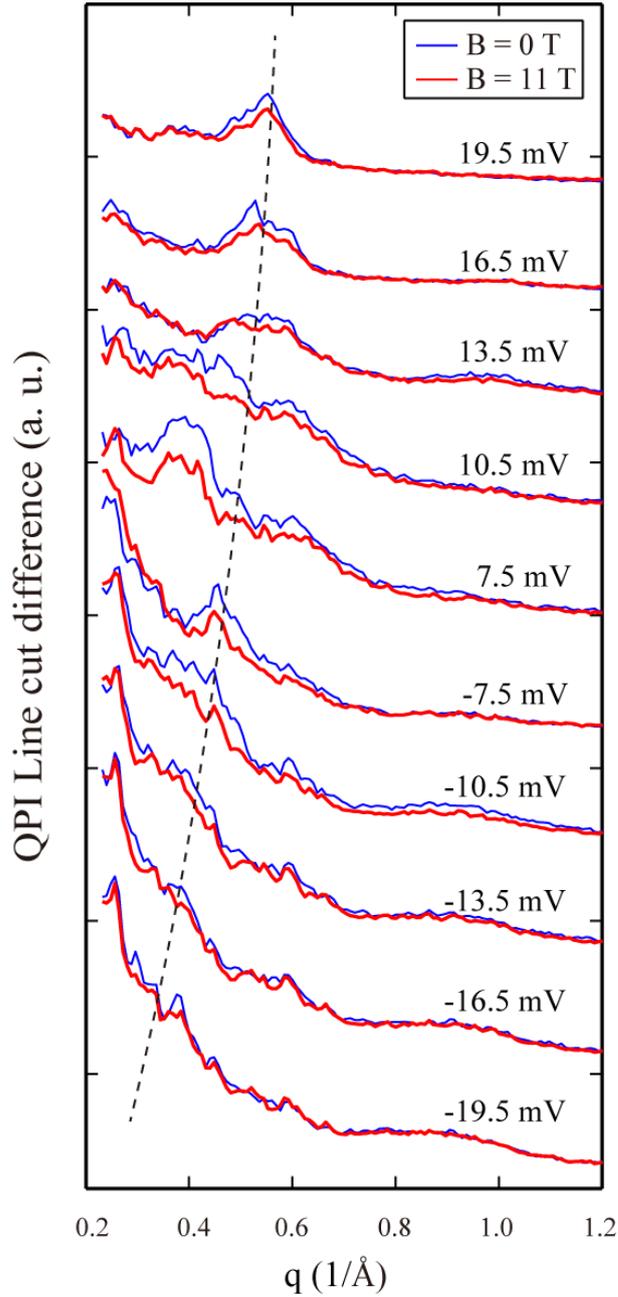

**Fig. s9 | Comparison of the FFT line cuts taken at B = 0T and B = 11T.** The line cut are taken along the yellow dashed line in Fig. 3b (which is (0,0) – (0, 2π) direction). The dashed curve follows the dispersion of Ring 1. The suppression of the Ring 1 at B=11T can be observed. Considering BZ folding, the Ring 1 should have three sub-sets (Fig. s6b), which cannot be clearly separated here. For $s_\pm$ wave pairing, the three sub-sets should behave differently under magnetic field, because two of them are sign-preserving and the other is sign-changing (see Fig. s7c). This behavior is not evident here.

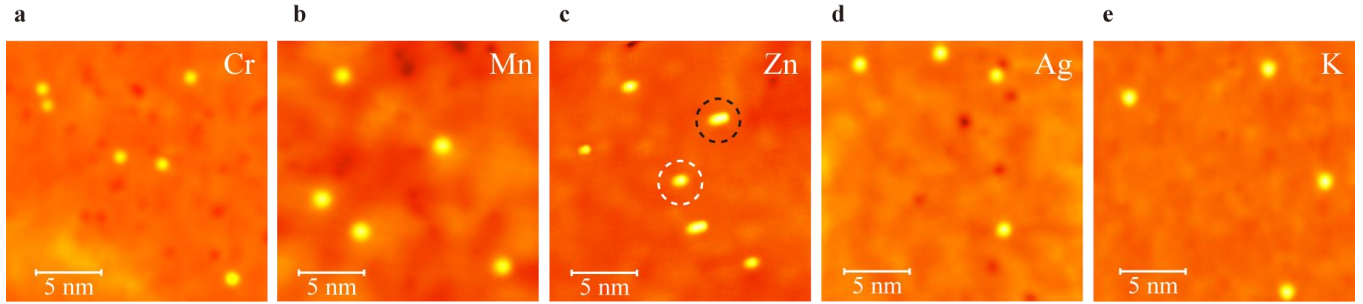

**Fig. s10 | (a-e):** Lager scale STM images of the impurity atoms deposited on single-layer FeSe/SrTiO$_3$. The bright circular protrusions on (a)~(e) are corresponding to Cr, Mn, Zn, Ag and K single adatoms, respectively. For Zn, sometimes dimmers can be found on the surface as marked by the black dashed circle in panel **c**, nevertheless the spectra shown in Fig. 5 are taken on a Zn monomer (marked by white dashed circle).